# A Global Cloud Map of the Nearest Known Brown Dwarf

I. J. M. Crossfield[1]*, B. Biller[1,2], J. E. Schlieder[1], N. R. Deacon[1], M. Bonnefoy[1,3], D. Homeier[4], F. Allard[4], E. Buenzli[1], Th. Henning[1], W. Brandner[1], B. Goldman[1], T. Kopytova[1,5]

Affiliations:

[1]Max Planck Institut für Astronomie, Königstuhl 17, 69117 Heidelberg, Germany

[2]Institute for Astronomy, University of Edinburgh, Blackford Hill, Edinburgh EH9 3HJ, UK

[3]UJF-Grenoble 1 / CNRS-INSU, Institut de Planétologie et d'Astrophysique de Grenoble (IPAG) UMR 5274, 38041 Grenoble, France

[4]CRAL-ENS, 46 Allée d'Italie, 69364 Lyon Cedex 07, France

[5]International Max-Planck Research School for Astronomy and Cosmic Physics at the University of Heidelberg, Königstuhl 17, 69117 Heidelberg, Germany

*Corresponding author:  ianc@mpia.de

**Brown dwarfs – interstellar bodies more massive than planets but not massive enough to initiate the sustained hydrogen fusion that powers self-luminous stars[1,2] – are born hot and slowly cool as they age.  As they cool below ~2300 K, liquid or crystalline particles composed of calcium aluminates, silicates, and/or iron condense into atmospheric "dust"[3,4] which disappears at still cooler temperatures (~1300 K)[5,6].  Models to explain this dust dispersal include both an abrupt sinking of the entire cloud deck into the deep, unobservable atmosphere[5,7] or breakup of the cloud into scattered patches[6,8] (as seen on Jupiter and Saturn[9]), but to date observations of brown dwarfs have been limited to globally integrated measurements[10]; such measurements can reveal surface inhomogeneities but cannot unambiguously resolve surface features[11].  Here we report a two-dimensional map of a brown dwarf's surface that allows identification of large-scale bright and dark features, indicative of patchy clouds. Geographic localization of such features, and the ability to create time-lapsed extrasolar weather movies in the near future, provide important new constraints on the formation, evolution, and dispersal of clouds in brown dwarf and extrasolar planet atmospheres.**

The recent discovery of the Luhman 16AB system (also called WISE J104915.57-531906.1AB; Ref. 12) revealed two brown dwarfs only 2 parsecs away, making these the closest objects to the Solar system after the alpha Centauri system and Barnard's star. Both of these newly-discovered brown dwarfs are near the dust clearing temperature[13,14], and one (Luhman 16B) exhibits strong temporal variability of its thermal radiation consistent with  a rotation period of 4.9 hr[15]. Luhman 16AB's proximity to Earth makes these the first substellar objects bright enough to be studied at high precision and high spectral resolution on short timescales, so we observed both of these brown dwarfs for five hours (one rotation period of Luhman 16B) using the CRIRES spectrograph[16] at ESO's Very Large Telescope to search for spectroscopic variability.

Absorption features from CO and $H_2O$ dominate the brown dwarfs' spectra, as shown in Fig. 1. The two objects have similar spectra but the absorption lines are broader for the B component: it exhibits a projected equatorial rotational velocity of 26.1 +/- 0.2 km/s, vs. 17.6 +/- 0.1 km/s for Luhman 16A. Taking Luhman 16B's rotation period[15] and considering that evolutionary models predict these objects to be 1.0+/- 0.2 times the radius of Jupiter[17], Luhman 16B's rotation axis must be inclined ≤30 deg from the plane of the sky; i.e., we are viewing this brown dwarf nearly equator-on. If the two brown dwarfs' axes are closely aligned (like those of the planets in our Solar system) then Luhman 16A rotates more slowly than Luhman 16B and the objects either formed with different initial angular momentum or experienced different accretion or spin-braking histories. Alternatively, if the two brown dwarfs have comparable rotation periods (as tentatively indicated by recent observations[18]) then the two components' rotation axes must be misaligned, which would imply either an initially aligned system (like the Solar system) which was subsequently perturbed or a primordial misalignment (in contrast to the close alignment more typically observed for pre-main sequence stellar binaries[19]). Measuring Luhman 16A's rotation period is the best way to determine whether the brown dwarfs' axes are currently aligned or misaligned.

Our data clearly show spectroscopic variability intrinsic to Luhman 16B, and this brown dwarf's rapid rotation allows us to produce the global surface map shown in Fig. 2 using Doppler Imaging techniques[20,21]. This produces a map that shows a large, dark, mid-latitude region; a brighter area on the opposite hemisphere located close to the pole; and mottling at equatorial latitudes.

A natural explanation for the features seen in our map of Luhman 16B is that we are directly mapping the patchy global clouds inferred to exist from observations of multiwavelength variability[15,18]. In this explanation, the dark areas of our map represent thicker clouds that obscure deeper, hotter parts of the atmosphere and present a higher-altitude (and thus colder) emissive surface, whereas bright regions correspond to holes in the upper cloud layers that provide a view of the hotter, deeper interior. This result is also consistent with previous suggestions of multiple stratified cloud layers in brown dwarf atmospheres[4,10,11]. Because our mapping is mostly sensitive to CO, the map could in principle show a combination of surface brightness (i.e., brightness temperature) and chemical abundance variations. Coupled models of global circulation and atmospheric chemistry[22], maps obtained via simultaneous observations of multiple molecular tracers, and/or simultaneous Doppler Imaging and broadband photometric monitoring could distinguish between these hypotheses.

The high-latitude bright spot could be similar to the polar vortices seen on Jupiter and Saturn and predicted to exist on highly irradiated gas giants in short-period orbits around other stars[23]; in this case, the high-latitude feature should still be visible in future maps of Luhman 16B. Jupiter and Saturn exhibit prominent circumplanetary banding, but (as described in the Methods section) our analysis is not sufficiently sensitive to detect banding on Luhman 16B. Furthermore, assuming a mean horizontal windspeed of ~300 m/s (as predicted by global circulation models of brown dwarfs at these temperatures[24]) the Rhines relation[25] predicts that Luhman 16B should exhibit roughly ten bands from pole to pole – too many to resolve with our 18 deg-wide map cells.

Long-term monitoring of Luhman 16B suggests that its weather conditions change rapidly but remain at least partly coherent from one night to the next[15], a result which indicates that the characteristic timescale for evolution of global weather patterns is of order one day. In this case, successive full nights of Doppler Imaging could observe the formation, evolution, and breakup of global weather patterns – the first time such a study is possible outside the Solar System. Such measurements would provide a revolutionary new benchmark against which to compare global circulation models of dusty atmospheres[24,26], and could perhaps even measure differential rotation in Luhman 16B's atmosphere[27]. Future mapping efforts should reveal whether we are mapping variations in temperature, cloud properties, or atmospheric abundances: high-resolution spectrographs with broader wavelength coverage than CRIRES should provide better sensitivity & spatial resolution[28], perhaps sufficient to search for banded cloud structures. Instruments with broader wavelength coverage will also allow maps to be made at multiple wavelengths and using independent molecular tracers (e.g. $H_2O$). In addition, a few other variable brown dwarfs may be bright enough for these technique to be applied. Although the day sides of hot, short-period gas giant planets can also be mapped using occultations under favorable conditions[29], model degeneracies may prevent these efforts from achieving a spatial resolution comparable to that achievable with Doppler Imaging[30]. Thus, Doppler Imaging in general, and Luhman 16B in particular, represent the best opportunity to challenge and improve our current understanding of the processes that dominate the atmospheres of brown dwarfs and of giant extrasolar planets.

**Methods Summary**

We extract and calibrate our spectroscopic data using standard techniques (Extended Data Figs. 1 and 2) and look for temporal changes in the mean spectral line profiles. Luhman 16B exhibits strong spectroscopic variability but we see no evidence for similar variations in our simultaneously-acquired observations of Luhman 16A (Extended Data Fig. 3). A simplified analysis using a parameterized spot model verifies that our observations are consistent with rotationally induced variations (Extended Data Fig. 4). We then produce our global map of brown dwarf Luhman 16B using Doppler Imaging.

The technique of Doppler Imaging relies on the varying Doppler shifts across the face of a rotating object and has been widely used to map the inhomogeneous surfaces of many rapidly rotating stars[20,21]. As darker regions rotate across the visible face of the brown dwarf, the Doppler-broadened absorption line profiles exhibit deviations at the projected radial velocities of the darker areas. Features near the equator cause changes across the entire line profile and move across the full span of velocities; features at higher latitudes move more slowly, experience smaller Doppler shifts, and affect a narrower range of velocities.

Our modeling framework is based on that described in Ref. 20. We break the brown dwarf's surface into a 10x20 grid, giving an effective equatorial cell size of roughly 20,000 km. The the recovered maps do not significantly change if we use finer resolution. We verify our analysis by constructing a number of Doppler Images using simulated data. These simulations demonstrate that we can robustly detect large, isolated features with strong brightness temperature contrasts (Extended Data Fig. 5) but that we are not sensitive to axially symmetric features such as zonal banding (Extended Data Fig. 6).

**Acknowledgments**  We thank Prof. A. Hatzes for advice on Doppler Imaging analysis, Dr. J. Smoker for helping to plan and execute our observations, and Dr. E. Mills for help designing figures. Based on Director's Discretionary observations made with ESO Telescopes at the Paranal Observatory under program ID 291.C-5006(A); data are available in the ESO Data Archive, and Fig. 1 is available as an electronic supplement to this work. D.H. has received support from the European Research Council under the European Community's Seventh Framework Programme (FP7/2007-2013 Grant Agreement no. 247060). M.B., D.H., and F.A. acknowledge support from the French National Research Agency (ANR) through project grant ANR10-BLANC0504-01. E.B. is supported by the Swiss National Science Foundation (SNSF). Atmosphere models have been computed on the Pôle Scientifique de Modélisation Numérique at the ENS de Lyon, and at the Gesellschaft für Wissenschaftliche Datenverarbeitung Göttingen in co-operation with the Institut für Astrophysik Göttingen. IRAF is distributed by the National Optical Astronomy Observatory, which is operated by the Association of Universities for Research in Astronomy (AURA) under cooperative agreement with the National Science Foundation. PyRAF is a product of the Space Telescope Science Institute, which is operated by AURA for NASA. We also thank contributors to SciPy, Matplotlib, AstroPy, and the Python Programming Language.

**Author contributions**  I.J.M.C. coordinated the project and conducted all analyses described herein. I.J.M.C., B.B., J.S., N.R.D., M.B., W.B., B.G., and T.K. assisted in obtaining the spectroscopic observations. I.J.M.C., B.B., J.S., N.D., M.B., D.H., F.A., E.B., Th.H., and W.B. contributed to the manuscript. J.S. provided an independent analysis of the projected rotational velocity and radial velocity of both brown dwarfs. B.B. and N.R.D. provided advice on binary dynamics. D.H. and F.A. provided advice on brown dwarf atmospheric processes, and the spectral models used for the data calibration, Least Squares Deconvolution, and Doppler Imaging.

**Author information**  Reprints and permissions information is available at www.nature.com/reprints. The authors have no competing financial interests. Correspondence should be addressed to I.J.M.C. (ianc@mpia.de).

**Figure Legends:**

**Figure 1**. High-resolution, near-infrared spectra of the Luhman 16AB brown dwarfs (black curves). Essentially all absorption features are real: the vertical ticks indicate absorption features in the brown dwarfs' spectra from $H_2O$ (blue) and CO (red), and residual features from the Earth's atmospheric absorption (gray). The lines of the B component are broader, indicating a higher projected rotational velocity: thus either the brown dwarfs' rotation axes are misaligned or Luhman 16B formed with or developed a shorter rotation period than its companion. The gaps in the spectra correspond to physical spaces between the four infrared array detectors. The plotted data represent the mean of all spectra, and are available as an electronic data supplement. Luhman 16A's spectrum has been offset vertically for clarity.

**Figure 2.** Surface map of brown dwarf Luhman 16B, which clearly depicts a bright near-polar region (seen in the upper-right panels) and a darker mid-latitude area (lower-left panels) consistent with large-scale cloud inhomogeneities. The lightest and darkest regions shown correspond to brightness variations of roughly ±10%. The time index of each projection is indicated near the center of the figure.

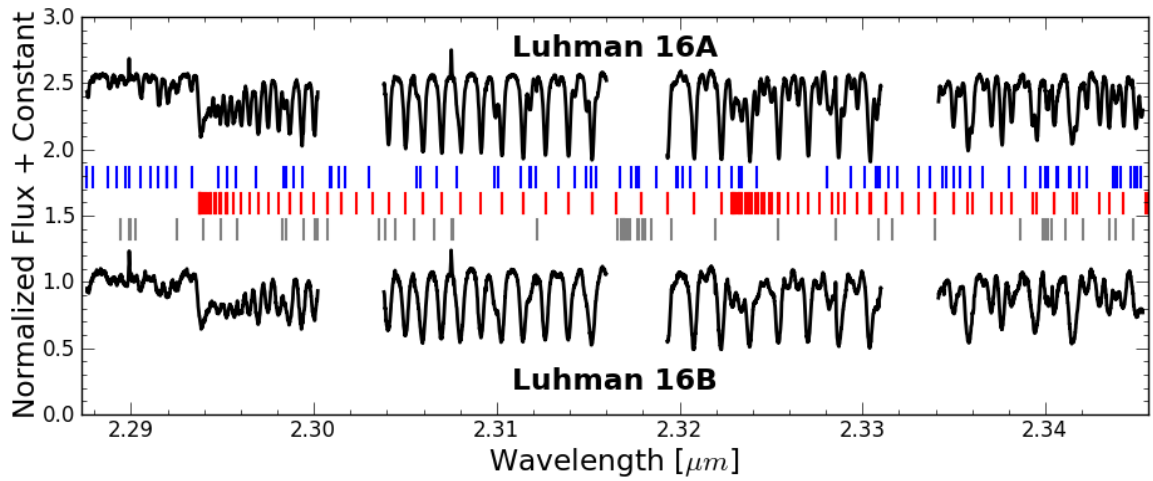
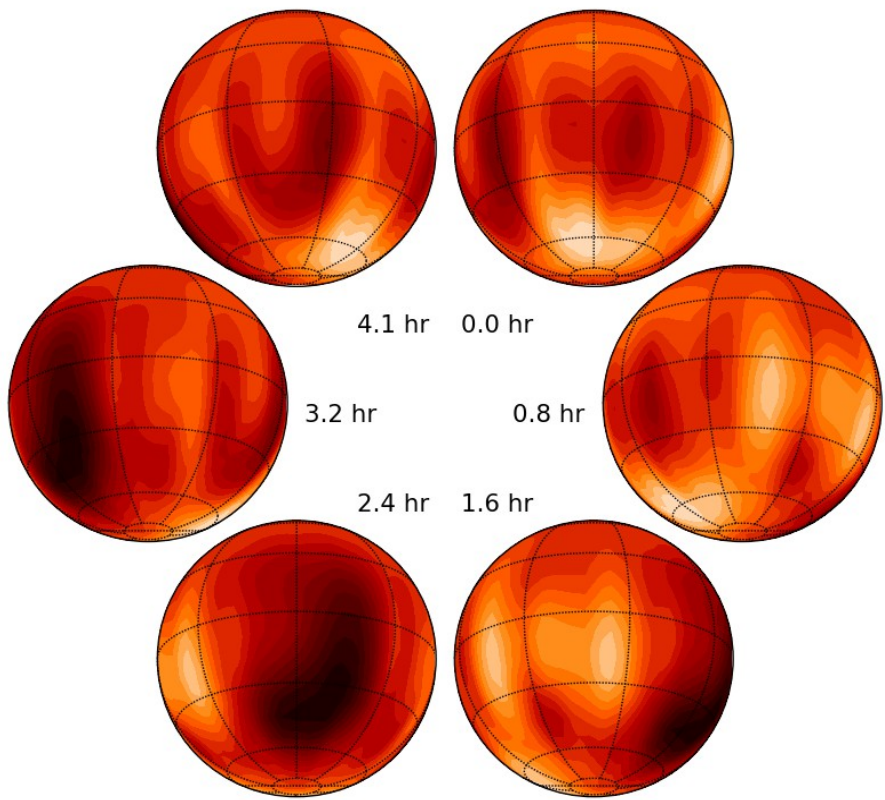

**Methods**

**Observation and Data Reduction:**

We observed the Luhman 16AB system for five hours with VLT/CRIRES[16] on UT 2013-05-05. Our spectra span wavelengths from 2.288—2.345 µm in order to cover the CO (3,1) and (2,0) bandheads. During our observations the spectrograph slit was aligned to the binary position angle, so that both brown dwarfs were observed contemporaneously. The telescope was nodded along the binary axis to subtract the emission from the infrared-bright sky using standard techniques. A small, random offset was applied to each nod position to mitigate bad detector pixels. Observing conditions were good: seeing was roughly 0.5", humidity was <10%, airmass ranged from 1.2—1.6, and the moon was down during our observations. We spatially resolved the two brown dwarfs and over five hours obtained 56 spectra with exposures of 300 s each. We calibrate the raw CRIRES data frames using the standard CRIRES esorex data reduction routines, combining spectra in sets of four to boost signal to noise. We extract one-dimensional spectra from both brown dwarfs in each combined frame using the standard astronomical IRAF data analysis package.

We use these extracted, raw spectral data to precisely measure both brown dwarfs' physical parameters. We follow past work using high-precision infrared spectroscopy and employ a forward-modeling approach[31,32] to calibrate our data. This analysis method transforms high-resolution spectra of the telluric transmission[33] and of model brown dwarf atmospheres[34] into a simulated CRIRES spectrum using an appropriate set of instrumental and astrophysical parameters. We use a model whose free parameters are the radial velocity of the model brown dwarf, rotational broadening ($v \sin i$) and linear limb-darkening coefficients[35], two multiplicative scaling factors for the telluric and brown dwarf models, polynomial coefficients that convert pixel number into wavelength, and coefficients for a low-order polynomial to normalize the continuum. As weights in the fits we use the uncertainties reported by IRAF after scaling these so that the weighted sum of the residuals equals the number of data points. The effect of all this is to place all observations on a common wavelength scale, to remove the effects of variable telluric absorption and spectrograph slit losses (from slight guiding errors or changes in seeing), and to estimate the astrophysical parameters listed above. Extended Data Fig. 1 shows examples of the raw and modeled data in this approach, and all calibrated spectra are shown in Extended Data Fig. 2. For each brown dwarf, we take as uncertainties the standard deviation on the mean of the measurements from each of the 14 spectra.

We apply the above modeling approach using a wide range of model brown dwarf spectra from the BT-Settl library[34], a set of models computed with the PHOENIX code that spans effective temperatures ($T_{eff}$) of 1000–1600 K and surface gravities ($\log_{10} g$) of 4.0—5.5. We perform a separate fit to data from each of the four CRIRES detectors and find that the same model does not always give the best fit to the data from all four detectors. Judging from the residuals to the fits (see Extended Data Fig. 1), this effect results from inaccuracies in both the adopted telluric spectrum and in the brown dwarf atmospheric models. Considering all these ambiguities, we find that the models with $\log_{10} g$ = 5.0 and $T_{eff}$=1500 and 1450 K (for components A and B, respectively) give the best fits to data from all four detectors. There is some degeneracy between temperature and surface gravity, with greater $T_{eff}$ allowing somewhat higher $\log_{10} g$. Brown dwarf atmospheres have never before been tested at this level of precision and so we do not interpolate between models to marginally improve the quality of the fit. The effective temperatures estimated from our analysis moderately exceed the values reported by previous studies[13,14], and we attribute this difference to the well-known phenomenon that the effective temperature estimated from fitting model spectra to the CO bandheads typically exceeds the temperature derived from integrating the broadband spectral energy distribution[36,37]. Comparison of future models to these data should be highly instructive in refining substellar atmospheric models. In the analyses that follow, we use the BT-Settl models with the parameters given above; using slightly different model parameters does not change our conclusions.

To properly conduct a Doppler Imaging analysis we must account for the radial velocity (RV) shift of the brown dwarfs. We measure RVs for the A and B components of 20.1 ± 0.5 km/s and 17.4 ± 0.5 km/s, respectively, relative to the Solar system barycenter; the uncertainties in these absolute measurements are dominated by systematic uncertainties in our instrument model. While the RVs of Luhman 16A exhibit little internal scatter during our observations, we see an anomalous deviation (lasting from 1.5 to 3 hr after the start of observations) of roughly 1 km/s in the RV measurements of Luhman 16B. Assuming that the systematic effects in measuring RVs are common to our observations of both brown dwarfs, and examining only the spectra taken outside the time of anomalous RVs, we obtain a relative RV between the components of 2800 ± 50 m/s. This measurement is consistent with the orbital velocity expected between two old brown dwarfs in a few-decade orbit[14], and indicates that it will eventually be possible to test brown dwarf evolutionary models[17] by measuring the individual component masses via a full 3D orbital solution using the system's RVs and astrometry[38,39].

To enhance our sensitivity, we use the technique of Least Squares Deconvolution[40] (LSD) to transform each spectrum into a single, high-S/N mean absorption line. Deviations in the resulting mean line profiles are difficult to see with the unaided eye, but after subtraction of the night's mean line profile variations are apparent. Extended Data Fig. 3 shows the resulting temporal evolution in the deviations from the global mean line profile: the rotational signature of Luhman 16B's inhomogeneous surface is clearly visible, dominated by rotation of a darker region into and then out of view. Hints of brighter regions are visible at other times. We find that the total absorption depth of the mean line profile decreases by ~4% during this period. No such coherent signatures are observable beyond Luhman 16B's projected rotational velocity of +/-26.1 km/s, and we do not see any such time-variable phenomena in our simultaneously-acquired spectra of Luhman 16A.

**Spot modeling:**
To interpret our LSD line profiles, we first implement a simple spot model similar to that used to interpret photometry of variable brown dwarfs[11]. This initial toy model assumes that Luhman 16B's surface is dominated by a single spot. We divide the surface into a grid, regularly spaced in latitude and longitude. A 10x20 grid (18 degrees across each cell) is sufficient for the analysis to converge. The spot is assumed to be circular and the remainder of the photosphere is assumed to have uniform surface brightness with a linear limb-darkening law. The free parameters are the brightness of the spot relative to the photosphere and the spot's radius, latitude, and longitude. For a given set of parameters we generate a surface map with the specified surface brightness distribution. For each grid cell we then use the projected visible area and apparent flux, and the cell's rotational Doppler shift, to generate a set of rotationally-broadened line profiles corresponding to the time of each observation. Each line profile is continuum-normalized, and the resulting set of simulated data is compared to the observed LSD line profiles.

To estimate the uncertainty on the spot parameters we use the emcee tool[41], which implements an affine-invariant Markov-Chain Monte Carlo approach. We initialize 150 chains near a set of reasonable guess parameters (final results are insensitive to this guess) and run all chains for 1500 steps. After this initial "burn-in" phase the Markov chains are randomized and have lost any memory of their initial starting conditions; we discard the initial steps and run the chains for an additional 1500 steps, afterward verifying that they are well-mixed both by examination of the autocorrelation of the individual chains and by visual inspection of the likelihood and parameter values of the chains. Extended Data Fig. 4 shows the resulting posterior distributions of the spot's latitude, radius, and surface brightness assuming $i$=30 deg; the results do not change significantly for smaller values of $i$. In this model the dark spot lies ≤31 deg from the equator, has a radius of 33 +/- 7 deg, and is 88+/-3% as bright as the surrounding photosphere. This result implies a photometric variation of ~3%, consistent with the wide range of variability seen from this system[15,18]. However, such parametrized models typically exhibit strong degeneracies and tend not to lead to unique maps of brown dwarfs' surface brightness distributions[11].

**Doppler Imaging:**
We construct our Doppler Imaging (DI) model as described by Ref. 20 using the same 10x20 grid and line profile simulation techniques described immediately above. Our results do not change significantly if we increase the model's spatial resolution. Instead of an arbitrarily-parametrized spot, in the DI model there are 200 free parameters: the contributions to the line profile from each grid cell. Because there are ~35 pixels across each of 14 mean line profiles, we nominally have 490 constraints; thus the problem appears well-posed and simple matrix techniques (e.g., Singular Value Decomposition) would seem to be sufficient. However, it has long been recognized that such an approach yields extremely noisy maps[20,21,42] often with nonphysical values (e.g., negative surface brightness in some cells). Regularization is needed, and we employ a maximum entropy approach[43] in which the merit function is $Q = \chi^2 - a\,S$ (where $\chi^2$ has its usual meaning, $S$ is the image entropy of the map's grid cells, and $a$ is a hyperparameter that determines the balance between goodness of fit and entropy). We minimize $Q$ using a standard multivariate optimizer, and we speed up convergence by calculating the analytical gradients of $Q$ relative to the brightnesses of the map cells.

Our data are noisier, and the spectroscopic variations weaker, than in typical DI analyses of stars, so we tune $a$ to minimize the appearance of spurious features while maintaining the fidelity of the resulting map. We do this by generating a number of synthetic surface maps, simulating their line profiles and adding Gaussian noise of the same amplitude as we find in our observed LSD line profiles, and minimizing $Q$ for various choices of $a$. An example of one such simulation and recovery is shown in Extended Data Fig. 5, using the same value of $a$ as in the analysis leading to Fig. 2. This analysis demonstrates that the prominent mid-latitude and polar features are likely real, while the lower-contrast equatorial features may be more affected by noise. The longitudinal elongation of equatorial features is a known feature of DI maps[44], so features near the equator may be narrower than they appear. The main features in our recovered map do not change for small variations in the DI modeling parameters. Finally, we find that although we cannot yet directly measure $i$ with the current data[45], our maps do not change much for expected values of $i$ (0-30 deg).

**Zonal Banding and Brown Dwarf Line Profiles:**
The detection of axisymmetric features (such as zonal banding) via Doppler Imaging is more challenging than the detection of features lacking such symmetry: the latter are easily seen via their time-variable effects on the line profiles, but the former can only be distinguished by discerning deviations of the mean line shape from the modeled profile. We ran a number of simulated DI analyses on brown dwarfs with various levels of banding. Extended Data Fig. 6 shows one such example , which is typical insofar as it demonstrates our inability to recover even strong, large-scale zonal bands. Even if the band contrast were 100%, our simulations show that recovery of such features would be only tentative given the current precision of our data. Future observations at higher precision should have greater sensitivity to such features, and these efforts will therefore become more susceptible to the spurious axisymmetric bands that can result from DI analyses performed with inappropriate line profile shapes[21,27,46]. Below, we therefore consider possible sources of uncertainty in modeling the detailed line shapes probed by our analysis.

In the PHOENIX synthetic atmosphere and spectral model employed in our analysis, the strongest molecular lines are modelled as regular, symmetric Voigt profiles extending out to a maximum half width of 20 cm$^{-1}$. Beyond this detuning, effects of asymmetry and mixing of neighbouring lines no longer allow an adequate representation of the wings by a simple Lorentzian. A generic half width at half maximum (HWHM) of 0.08 cm$^{-1}$ bar$^{-1}$ at 296 K and a temperature exponent of 0.5 was assumed for the Lorentz (pressure broadening) part of all molecular lines[34], and Doppler broadening is calculated for the thermal velocity plus an isotropic microturbulence of 0.8 km s$^{-1}$. The part of the spectrum covered by our observations forms mainly at pressure levels of 1-2 bar[18]. Specifically, the wings of the strongest CO lines would form as deep as 4 bar, while the most central portion of the line cores form as high as the 10 mbar level. The corresponding atmospheric temperatures in these layers (1000-1500 K) yield a total Doppler broadening of the order of 1.1-1.25 km s$^{-1}$, while the half width of the collisional profile ranges from a few 100 m s$^{-1}$ in the cores up to nearly 10 km s$^{-1}$ in the outer wings.

The true line profiles might deviate in several respects from the assumptions used in the PHOENIX model. Microturbulence, which in stellar atmosphere modelling simply denotes a random Gaussian velocity distribution on scales small compared to the photon mean free path, has not been constrained very tightly for brown dwarfs yet. An ill-estimated microturbulence, particularly in the case of an anisotropic distribution with a stronger horizontal component, may affect the retrieval of surface features[21,27,46].

Radiation Hydrodynamic Simulations do allow us some insight into the dynamic structure of brown dwarf atmospheres, predicting horizontal RMS velocities of the order 0.3 km s$^{-1}$ for our case, compared to 3-5 times smaller values for the vertical component[47]. However, unlike in the case of typical stars mapped by DI in our case the total broadening is always dominated by the thermal velocity; so given the PHOENIX models' constant microturbulent velocity of 0.8 km s$^{-1}$, any realistic changes are unlikely to have noticeable impact on the line shapes. Pressure broadening of molecular lines, in contrast, is only poorly studied for stellar and substellar atmosphere conditions, i.e. for temperatures of 1000 K and higher and with molecular hydrogen and helium as main perturbers. Measurements of the broadening of CO lines in the fundamental band at 4.6 micron by noble gases and various other perturbers have yielded a HWHM of ca. 0.07 cm$^{-1}$ bar$^{-1}$ at 296 K for $H_2$ (Ref. 48). A study of the overtone band at 2.3 micron perturbed by various noble gases showed very similar widths to those in the fundamental[49], so it may be safe to assume that the $H_2$ broadening in this band is also comparable, and only about 12% smaller than our model value. The temperature dependence however could be stronger with a possible temperature exponent of 0.5-0.75 (Refs. 50, 51). In combination of all these effects we may expect the actual damping widths to be up to a factor of 2 smaller than assumed in our model. On the other hand, the actual atmospheric conditions also remain poorly constrained without a detailed spectral analysis or tighter limits on age and mass of the system. For an older and more massive brown dwarf an up to three times higher surface gravity with correspondingly larger atmospheric pressures is possible, which would affect the collisional damping part of the line wings, but not the Doppler cores. Finally, collisional perturbations are also known to shift molecular lines. This effect, and in particular its temperature dependence, is even less well studied for perturbers other than $H_2$ and He[49,52], but the shifts should be around an order of magnitude smaller than the HWHM and thus have little effect on the position of the line cores.

In conclusion, it seems feasible that future observations at higher precision could determine whether Luhman 16B exhibits zonal banding. At present, our current data are not sufficiently sensitive to address this issue.

**Extended Data Figure Legends:**

**Extended Data Figure 1:** Spectral calibration for Luhman 16A (top) and B (bottom). In the upper panels, the red curves show the modeled spectra, which mostly overlap the observed spectra (black). The lower panels show that the residuals to the fits are generally a few percent, with larger deviations apparent near CO bandheads (e.g., 2.294 and 2.323 μm) and strong telluric absorption lines (e.g., 2.290 and 2.340 μm).

**Extended Data Figure 2**: Same as Fig. 1, but showing the individual calibrated spectra of Luhman 16A (top) and B (bottom). The time of each observation is indicated at left.

**Extended Data Figure 3**: Luhman 16B shows strong rotationally-induced variability (right) while Luhman 16A does not (left). The color scale indicates the deviations from a uniform line profile as measured relative to the line continuum. Luhman 16B's variations are dominated by a dark region (diagonal streak, corresponding to a decrease of roughly 4% in equivalent width) that comes into view at 1.5 hr heading toward the observer, rotates across the brown dwarf to the receding side, and is again

hidden behind the limb at 3 hr. Brighter regions are visible at earlier and later times, but are less prominent. No significant spectroscopic variability is apparent for Luhman 16A, and no coherent features are seen beyond Luhman 16B's projected rotational velocity (vertical dashed lines). All these points indicate that we are detecting intrinsic spectroscopic variability from Luhman 16B.

**Extended Data Figure 4**: Posterior parameter distributions from our single-spot toy model, showing a large mid-latitude spot. The inner and outer curves in each panel indicate the 68.3% and 95.4% confidence regions. The plot shown assumes $i$=30 deg; smaller inclinations result in a slightly more equatorial spot, but the best-fit values do not significantly change.

**Extended Data Figure 5**: Simulated brown dwarf with spots, and the map recovered from Doppler Imaging. *Left*: Simulated variable brown dwarf seen at an inclination of $i$=30 deg. The dark and light spots are, respectively, 40% darker and 10% brighter than the photosphere; the dark streak is 10% darker and the polar spot is 20% brighter. *Right*: Surface map recovered from Doppler Imaging assuming noise levels similar to that seen in our observed data, after tuning the hyperparameter *a* to minimize spurious features. High-contrast features are recovered: the dark spot is in the correct location and the polar spot is only moderately distorted. The equatorial bright spot is visible in the recovered map, but it cannot be reliably distinguished from image artifacts that preferentially cluster near the equator. The dark stripe is not recovered. Thus our analysis can accurately recover strong features, but data quality precludes us from discerning smaller or fainter features.

**Extended Data Figure 6**: Simulated brown dwarf with spots and bands, and the map recovered from Doppler Imaging. *Left*: Simulated variable brown dwarf with the same surface features as in Extended Data Figure 5, but now also exhibiting zonal bands with an amplitude of +/-20% of the mean photospheric brightness level. *Right*: Surface map recovered from Doppler Imaging under the same assumptions as in Extended Data Figure 5. High-contrast, non-axisymmetric features are recovered as before, but we cannot recover even prominent global bands with the current precision of our data.

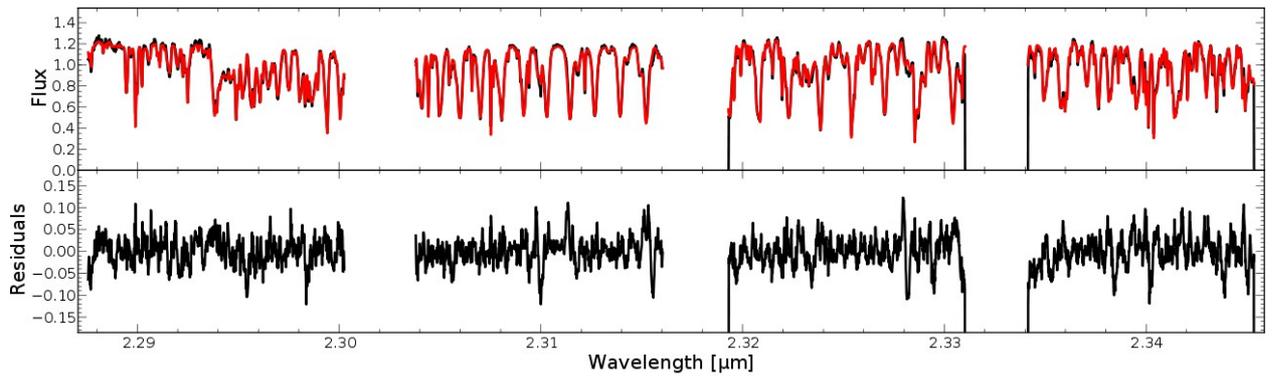
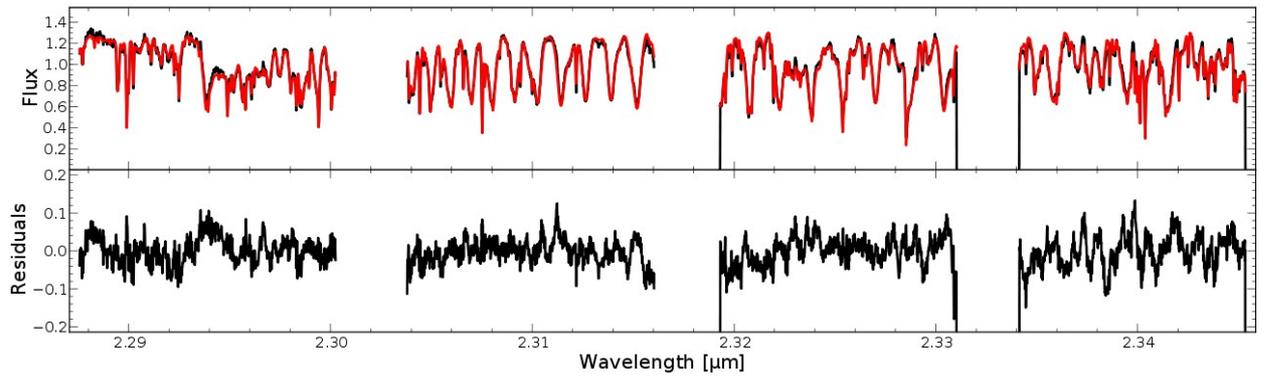
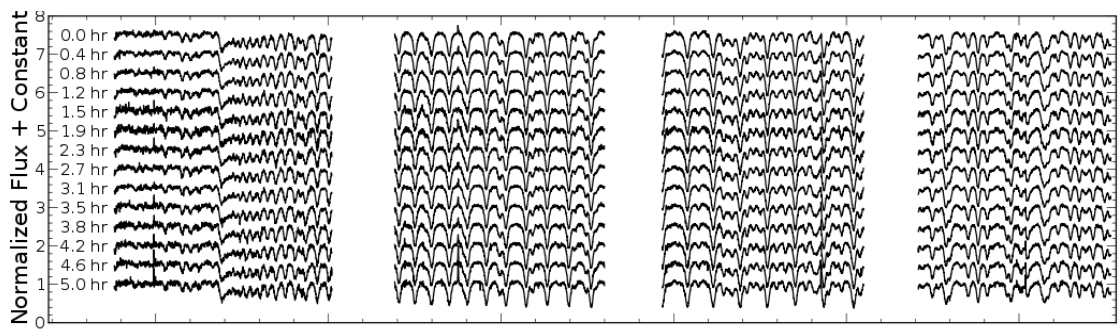
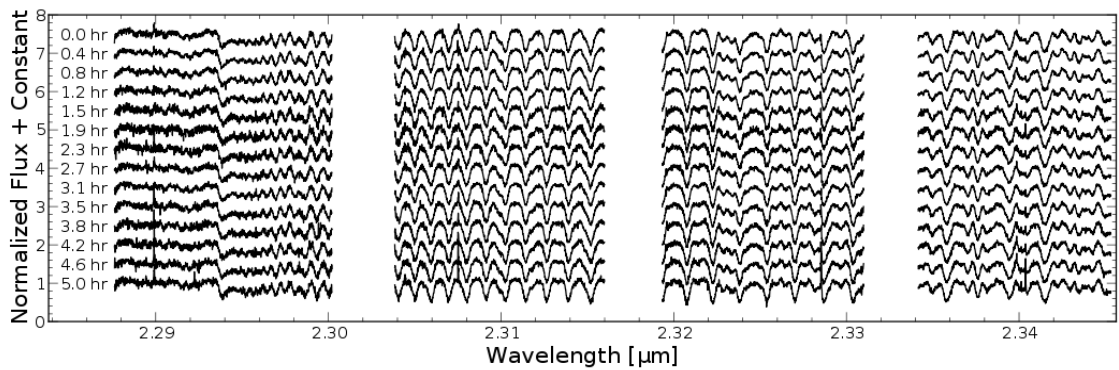
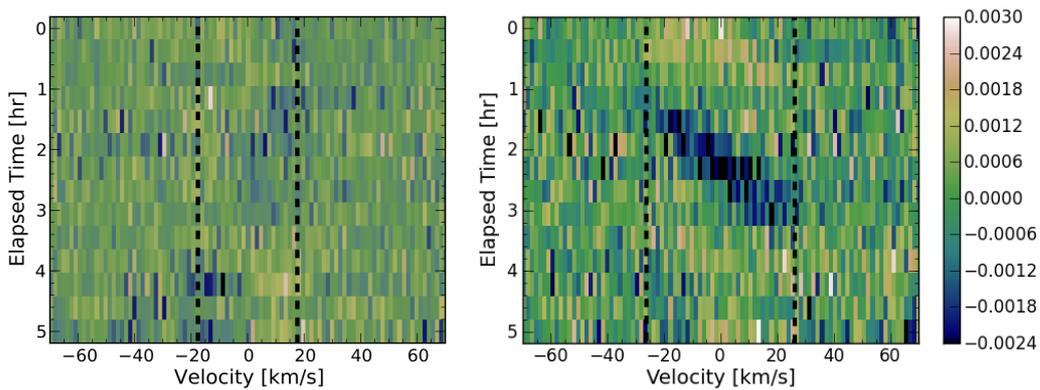

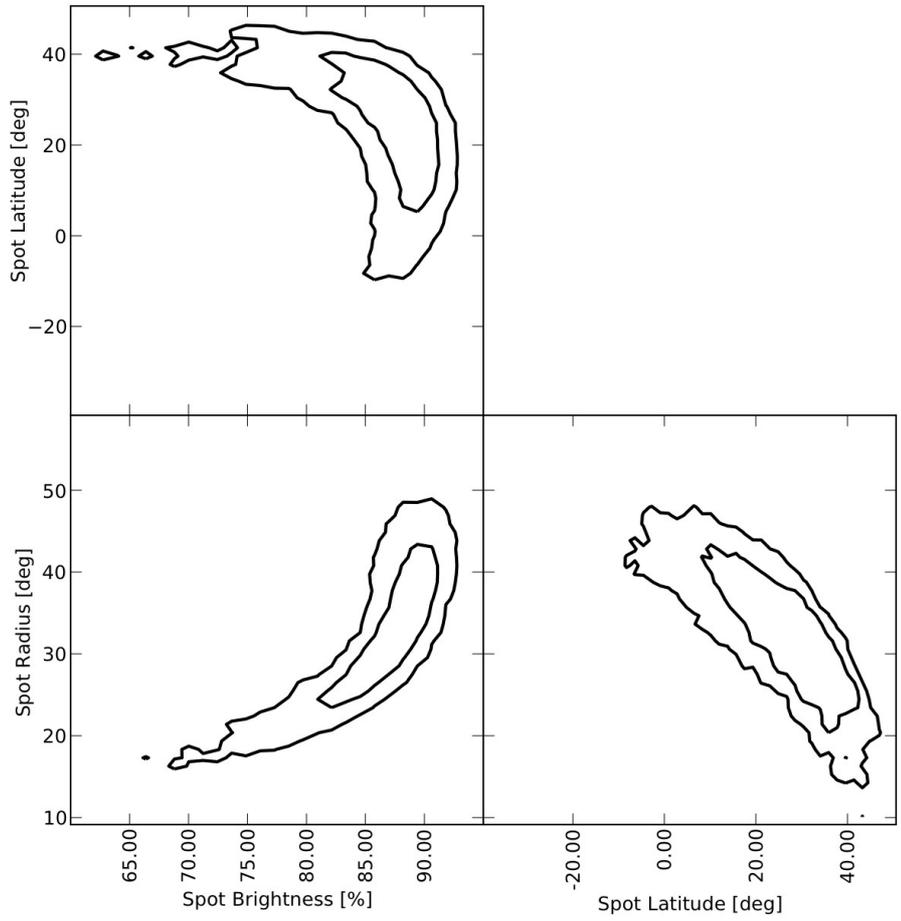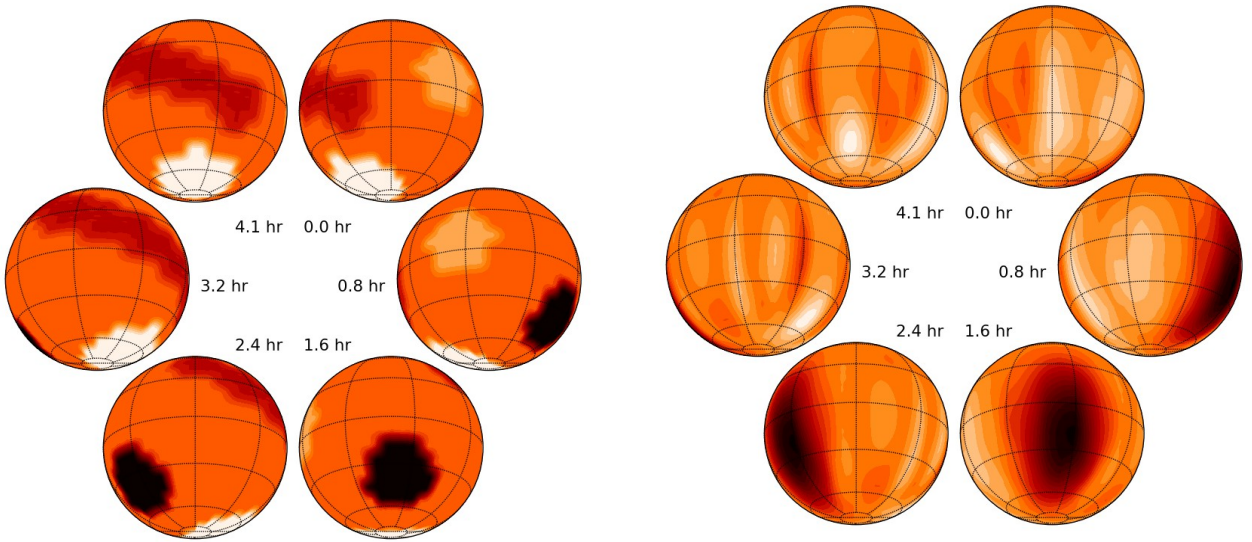

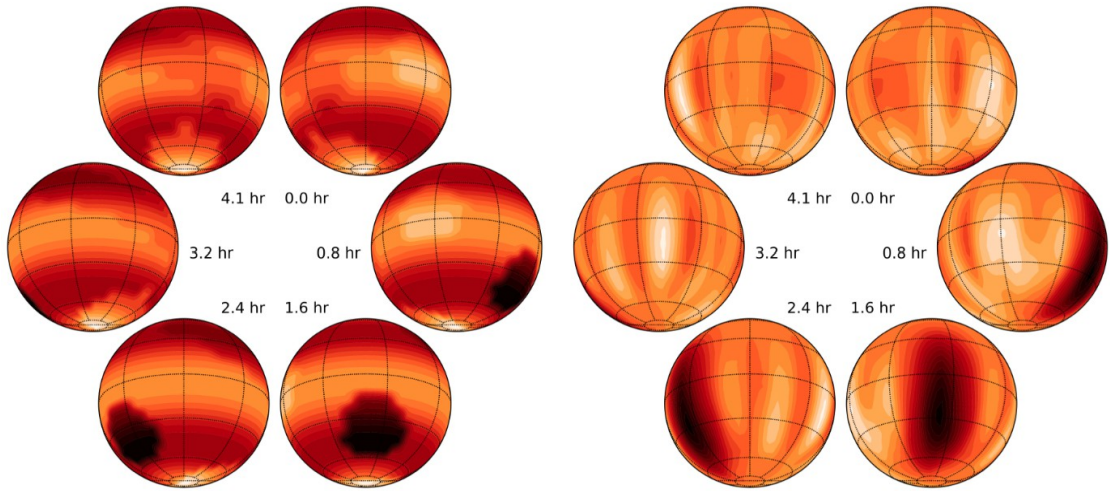